\documentstyle[12pt]{article}
\begin{document}
\begin{flushright}
hep-th/0506109\\
ICTP/June/2005
\end{flushright}
\vskip 2.5cm
\begin{center}
{\bf \Large {Nilpotent symmetries for a spinning relativistic 
particle\\ in augmented superfield formalism}}

\vskip 2.5cm

{\bf R.P.Malik}
\footnote{ Permanent Postal
Address: S. N. Bose National Centre for Basic Sciences,
Block-JD, Sector-III, Salt Lake, Kolkata-700 098, India.
Electronic Mail Address: malik@boson.bose.res.in.  }\\
{\it The Abdus Salam International Centre for Theoretical Physics,} \\
{\it Strada Costiera 11, 34014 Trieste, Italy} \\

\vskip 2cm

\end{center}

\noindent
{\bf Abstract}: 
The local, covariant, continuous, anticommuting and nilpotent
Becchi-Rouet-Stora-Tyutin (BRST) and anti-BRST symmetry transformations
for {\it all} the fields of a $(0 + 1)$-dimensional spinning relativistic
particle are obtained in the framework of 
augmented superfield approach to BRST
formalism. The trajectory of this super-particle, parametrized by a
monotonically increasing evolution parameter $\tau$, is embedded in a 
$D$-dimensional flat Minkowski spacetime manifold. This physically
useful one-dimensional system is considered on a three 
$(1 + 2)$-dimensional supermanifold which is parametrized by an even element
($\tau$) and a couple of odd elements ($\theta$ and $\bar\theta$) of the
Grassmann algebra. Two anticommuting sets of (anti-)BRST symmetry
transformations, corresponding to the underlying
(super)gauge symmetries for the above system, 
are derived in the framework of  
augmented superfield formulation where (i) the horizontality
condition, and (ii) the invariance of conserved quantities on the
supermanifold, play decisive roles. Geometrical interpretations
for the above nilpotent symmetries (and their generators)
are provided.

\baselineskip=16pt

\vskip .7cm



\newpage

\noindent
{\bf 1 Introduction}\\

\noindent
For the covariant canonical quantization of gauge theories
\footnote{These theories are endowed with the first-class constraints
in the language of Dirac's prescription for the classification of constraints.
The local (non-)Abelian 1-form interacting gauge theories provide
an almost exact theoretical basis for the three (out of four) fundamental 
interactions of nature.}, one of the most
elegant and intuitive approaches is the Becchi-Rouet-Stora-Tyutin (BRST)
formalism [1,2]. In this formalism, the unitarity and ``quantum'' 
gauge (i.e. BRST) invariance are very naturally respected at
any arbitrary order of perturbative computations for any arbitrary  physical
process allowed by the interacting gauge theories (where there
exists self-interaction as well as the coupling between the (non-)Abelian 
gauge field and the matter fields). In fact, the whole strength of 
BRST formalism appears in its full blaze of glory in the context of 
an interacting non-Abelian
gauge theory where the (anti-)ghost fields are required in the precise
proof of unitarity. To be more accurate, for every gluon loop (Feynman)
diagram, one requires a ghost loop diagram so that unitarity of the theory
could be maintained at any given order of perturbative calculation
(see, e.g., [3] for details).
In modern context, the BRST formalism is indispensable
in the realm of topological field theories [4-6], topological
string theories [7], string field theories [8], etc. There are well-known
connections of this formalism with the mathematics of differential geometry
and supersymmetries.

In our present endeavour, we shall be concentrating on the geometrical
aspects of the relationship between the BRST formalism and the
superfield formalism. To be more elaborate on this topic, 
it should be noted that, in the framework of usual
superfield formulation [9-14] of the BRST approach to $D$-dimensional 
$p$-form (with $p = 1, 2,.....$) Abelian gauge
theories, the gauge theory is considered first on a $(D + 2)$-dimensional 
supermanifold
%\footnote{We follow here the notations and conventions adopted in [12].
%The notation used, now a days, to denote the above six dimensional manifold
%is $(D, 2)$.}
 parametrized by $D$-number of even (commuting)
spacetime $x_\mu$ variables (with $\mu = 0, 1, 2...D-1)$
and a couple of odd (anticommuting) Grassmannian 
variables $\theta$ and $\bar\theta$ (with $\theta^2 = \bar\theta^2 = 0,
\theta \bar\theta + \bar\theta \theta = 0$). Then, the $(p + 1)$-form
super curvature $ \tilde F^{(p + 1)} = \tilde d \tilde A^{(p)}$ is 
constructed from the super exterior
derivative $\tilde d = dx^\mu \partial_\mu + d \theta \partial_{\theta}
+ d \bar\theta \partial_{\bar\theta}$ 
(with $\tilde d^2 = 0$) and the super $p$-form
connection $\tilde A^{(p)}$ defined on the $(D + 2)$-dimensional
supermanifold. This is subsequently equated, 
due to the so-called horizontality
condition [9-14], with the ordinary curvature
$(p + 1)$-form $ F^{(p + 1)} = d A^{(p)}$ defined
on the $D$-dimensional ordinary spacetime manifold with the exterior
derivative $ d = dx^\mu \partial_\mu$ (with $ d^2 = 0$)
 and the ordinary $p$-form connection $A^{(p)}$. The above
horizontality condition 
\footnote{For the 1-form non-Abelian gauge theory, the horizontality
condition $\tilde F^{(2)} = F^{(2)}$, where  2-form super curvature
$\tilde F^{(2)} = \tilde d \tilde A^{(1)} 
+ \tilde A^{(1)} \wedge \tilde A^{(1)}$
and 2-form ordinary curvature
$F^{(2)} = d A^{(1)} + A^{(1)} \wedge A^{(1)}$, leads to the
exact derivation of the nilpotent and anticommuting
(anti-)BRST symmetry transformations
for the non-Abelian gauge field and the corresponding (anti-)ghost fields
of the theory (see, e.g. [12] for details).}
is christened as the soul-flatness condition in [15]
which amounts to setting equal to zero all the Grassmannian components
of the (anti-)symmetric super curvature tensor that defines the
$(p + 1)$-form super curvature.
We shall be following, in our entire text, the notation adopted in [12] to
denote the above general supermanifold as $(D + 2)$. It should be noted,
however, that the modern notation 
is $(D, 2)$ for the same.

The process of reduction of the $(p + 1)$-form
super curvature to the ordinary $(p + 1)$-form curvature
due to the horizontality condition
(i) generates the nilpotent 
and anticommuting (anti-)BRST transformations for {\it only}
the gauge fields and the (anti-)ghost fields, (ii) provides the geometrical
interpretation for the nilpotent (anti-)BRST charges as the translational
generators $(\mbox {Lim}_{\bar\theta \to 0} (\partial/\partial\theta))
\mbox {Lim}_{\theta \to 0} (\partial/\partial\bar\theta)$ along the
$(\theta)\bar\theta$-directions of the $(D + 2)$-dimensional supermanifold,
(iii) leads to the geometrical interpretation for the nilpotency
property as the two successive translations 
(i.e. $(\partial/\partial\theta)^2 = (\partial/\partial\bar\theta)^2 = 0$)
along either of the Grassmannian directions, and
(iv) captures the anticommutativity of the nilpotent (anti-)BRST charges
(and the transformations they generate) in the relationship
$(\partial/\partial\theta) (\partial/\partial\bar\theta) +
(\partial/\partial\bar\theta) (\partial/\partial\theta) = 0$. It should 
be re-emphasized, however, that
all these nice geometrical connections between the BRST formalism
and the usual superfield formalism [9-14] 
are confined {\it only} to the gauge fields and 
the (anti-)ghost fields of a BRST invariant Lagrangian density of
the $D$-dimensional interacting
$p$-form Abelian gauge theory. The matter fields of an 
{\it interacting} $D$-dimensional $p$-form Abelian
gauge theory remain untouched
in the above superfield formalism as far as their nilpotent
and anticommuting (anti-)BRST symmetry transformations are concerned.

The above constraint due to the horizontality condition
(where  only $\tilde d$ and $d$ play important roles) has been
generalized to the constraints that emerge from
the full use of super ($\tilde d, \tilde \delta, \tilde \Delta$)
and ordinary ($d ,\delta, \Delta$) de Rham cohomological operators
(see, e.g., [16-20] for details). These complete set of restrictions 
on the supermanifold
lead to the existence of (anti-)BRST, (anti-)co-BRST 
and a bosonic (which is equal to the anticommutator of the (anti-)BRST
and the (anti-)co-BRST) symmetry transformations {\it together} for the
$(1 + 1)$-dimensional non-interacting 1-form (non-)Abelian
gauge theories. In the Lagrangian formulation,
the above kind of symmetries have also been shown to exist
for the $(3 + 1)$-dimensional
free Abelian 2-form gauge theory [21,22]. There exists a discrete
symmetry transformation for the above field theoretical models 
(in the Lagrangian formulation) which
corresponds to the Hodge duality $*$ operation of differential
geometry. Thus, the above models do provide a tractable 
set of field theoretical
examples for the Hodge theory. It is worthwhile to pinpoint, however,
that even the above new attempts of the superfield formalism (with the
full set of cohomological operators)
do {\it not} shed any light on the nilpotent symmetry transformations
associated with the matter fields.

In a set of recent papers [23-27], the above usual superfield formalism 
(with the theoretical arsenal of horizontality condition and its
generalizations) has
been augmented to include the invariance of the conserved currents and/or
charges on the supermanifolds. The latter constraints, on the
supermanifolds, lead to the derivation of the nilpotent and anticommuting 
(anti-)BRST symmetry transformations for the matter fields of the
interacting four dimensional 1-form (non-)Abelian gauge theories.
We christen this extended version of the superfield formalism
as the {\it augmented} superfield formulation
applied to the four dimensional interacting 1-form (non-)Abelian
gauge theories 
described by the corresponding (anti-)BRST invariant Lagrangian densities. 
It is worth emphasizing  that, in the framework
of augmented superfield formalism, all the geometrical interpretations,
listed in the previous paragraph, remain intact. As a consequence,
there is a very nice mutual consistency and complementarity between 
the old constraint (i.e. the horizontality condition)
and new constraint(s) on the supermanifolds. We do obtain,
as a bonus and by-product, all the nilpotent 
(anti-)BRST transformations for
{\it all} the fields (i.e. gauge fields, (anti-)ghost fields and matter fields) 
of an interacting 1-form (non-)Abelian gauge theory.

The purpose of the present paper is to derive the nilpotent (anti-)BRST 
transformations for {\it all} the fields, present in the description of
a free spinning relativistic particle (moving on a super world-line) in the 
framework of  augmented superfield formulation [23-27]. Our present 
endeavour is essential primarily
on four  counts. First and foremost, this formalism is being applied
to a supersymmetric system for the first time. It is worth pointing out that 
its non-supersymmetric counterpart
(i.e. the system of a free scalar relativistic particle) 
has already been discussed in the framework
of augmented superfield formulation in our earlier work [27]. Second,
to check the mutual consistency and
complementarity between (i) the horizontality condition, and (ii) the
invariance of conserved quantities on the supermanifold
for this physical system. These were found to be true in the case of (i)
a free scalar relativistic particle [27], (ii) the
interacting (non-)Abelian gauge theories in two 
$(1 + 1)$-dimensions (2D) [23,24],
(iii) the interacting (non-)Abelian gauge theories in
four $(3 + 1)$-dimensions (4D) of spacetime [25,26]. Third, to 
generalize our earlier
works [23-27], which were connected {\it only} with the gauge symmetries
and reparametrization symmetries, to the case where the supergauge symmetry 
also exists for the present system under discussion.
Finally, to tap the potential and power
of the above restrictions in the derivation of the nilpotent 
symmetries for the case of a new system 
where the fermionic as well as bosonic (i) the gauge fields (i.e. $\chi, e)$, 
and (ii) the (anti-)ghost fields (i.e. 
$(\bar c)c$) and $(\bar\beta)\beta$) do
exist in the Lagrangian description of  this 
{\it supersymmetric} system (cf. (2.7) below).

The contents of our present paper are organized as follows. In Sec. 2,
we very clearly discuss the essentials of the reparametrization,
gauge and supergauge symmetry transformations for the spinning massive
relativistic particle in the Lagrangian formulation. Two sets of anticommuting 
BRST symmetry transformations, that exist for the above system
under a very specific limit,  are also discussed in this section. 
Sections 3 and 4 are the centrals of our paper. Sec. 3 is devoted
to the derivation of the nilpotent (anti-)BRST symmetry transformations
(corresponding to the gauge symmetry transformations) in the framework of 
augmented superfield formalism. In the forthcoming section (i.e. Sec. 4), 
for the first time, we extend the idea of augmented superfield formalism
to obtain the (anti-)BRST symmetry transformations (corresponding to
the supergauge symmetry transformations) that
exist for the spinning relativistic particle. Finally, in Sec. 5, we
make some concluding remarks and point out a few future directions
for further investigations.\\

\noindent
{\bf 2 Preliminary: Nilpotent BRST Symmetries }\\

\noindent
Let us begin with the various equivalent forms of the 
reparametrization
invariant Lagrangians for the description of
a free massive spinning relativistic particle moving 
on a super world-line that is embedded in a $D$-dimensional
flat Minkowski target spacetime manifold.  These, 
triplets of appropriate Lagrangians, are [28,29]:
$$
\begin{array}{lcl}
&&L^{(m)}_{0} = m \; [\;(\dot x_\mu + i \chi \psi_\mu)^2\;]^{1/2}\; 
+ {\displaystyle \frac{i}{2} \; \Bigl (\psi_\mu\; \dot \psi^\mu
- \psi_5 \;\dot \psi_5 \Bigr ) - i \;\chi\; \psi_5\; m}, \nonumber\\
&& L^{(m)}_{f} = p_\mu \dot x^\mu - 
{\displaystyle \frac{1}{2}}\;e\; (p^2 - m^2)\; 
+ {\displaystyle \frac{i}{2} \; \Bigl (\psi_\mu\; \dot \psi^\mu
- \psi_5 \;\dot \psi_5 \Bigr )  
+ i\; \chi\; \Bigl (\psi_\mu p^\mu - \psi_5\; m \Bigr )}, \nonumber\\
&& L^{(m)}_{s} = 
{\displaystyle \frac{1}{2}\; e^{-1}\; \Bigl (\dot x_\mu 
+ i \chi \psi_\mu \Bigr )^2 + \frac{1}{2}}\; e \; m^2
+ {\displaystyle \frac{i}{2} \; \Bigl ( \psi_\mu \dot \psi^\mu
- \psi_5 \;\dot \psi_5 \Bigr ) - i\; \chi\; \psi_5 \;m}.
\end{array} \eqno(2.1)
$$
In the above, the mass-shell condition ($p^2 - m^2 = 0$),
the constraint condition $p \cdot \psi - m \psi_5 = 0$ and the force free
(i.e. $\dot p_\mu = 0$) motion of the {\it spinning} relativistic particle
are some of the key common 
features for (i) the Lagrangian with the square root $L^{(m)}_{0}$, (ii) the
first-order Lagrangian $L^{(m)}_f$, and (iii)
the second-order Lagrangian $L^{(m)}_s$. The constraints 
$p^2 - m^2 \approx 0$ and $p \cdot \psi - m \psi_5 \approx 0$ 
in $L_f^{(m)}$ are taken care of by the Lagrange
multiplier fields  $e (\tau)$ and $\chi (\tau)$ (with $\chi^2 = 0$) which
are (i) the bosonic and fermionic gauge fields  of the present system, 
respectively, and (ii) the analogues of the vierbein and Rarita-Schwinger 
(gravitino) fields
in the language of the supergravity theories. The Lorentz vector
fermionic fields $\psi_\mu (\tau)$ (with $\mu = 0, 1, 2....D-1$) are
the superpartner of the target space coordinate variable $x_\mu (\tau)$
(with $\mu = 0, 1, 2....D-1$) and classically they present spin
degrees of freedom. Furthermore, they anticommute with
themselves (i.e. $\psi_\mu \psi_\nu + \psi_\nu \psi_\mu = 0$)
and other fermionic field 
variables (i.e. $\psi_\mu \psi_5 + \psi_5 \psi_\mu = 0,
\psi_\mu \chi + \chi \psi_\mu = 0$) of the system under consideration. 
The  $\tau$-independent mass parameter
$m$ (i.e. the analogue of the cosmological constant term) is introduced  in 
our present system through the anticommuting (i.e. $\psi_5 \chi + \chi \psi_5
= 0, (\psi_5)^2 = - 1$, etc.) Lorentz scalar field $\psi_5 (\tau)$. The
momenta $p_\mu (\tau)$ (with $\mu = 0, 1, 2....D-1$), present 
in $L_f^{(m)}$, are canonically conjugate to the target space 
coordinate variable $x^\mu (\tau)$. 
It is evident that,
except for the mass parameter $m$, the rest of the field variables are
the functions of monotonically increasing
parameter $\tau$ that characterizes the trajectory (i.e. the super world-line)
of the massive spinning
relativistic particle. Here $\dot x^\mu = (d x^\mu/d \tau)
= e p^\mu - i \chi \psi^\mu$, $\dot \psi_\mu = (d \psi_\mu/ d\tau) = \chi p_\mu,
\dot \psi_5 = (d \psi_5/d \tau) = \chi m$ are the
generalized versions of 
``velocities'' of the massive spinning relativistic particle.

In what follows, we shall focus on the first-order Lagrangian $L_f^{(m)}$
for the discussion of the symmetry properties of the system. 
This is due to the fact that this Lagrangian is
comparatively simpler in the sense that there are no square roots
and there are no field variables in the denominator. Furthermore, it
is endowed with the maximum number of field variables and, therefore, is
interesting from the point of view of theoretical discussions. Under
an infinitesimal version of the reparametrization transformations
$\tau \to \tau^\prime = \tau - \epsilon (\tau)$, where $\epsilon (\tau)$
is an infinitesimal parameter, the field variables of $L_f^{(m)}$ transform
as
$$
\begin{array}{lcl}
\delta_r x_\mu &=& \epsilon \;\dot x_\mu,\; \;\;\qquad\;
\delta_r p_\mu = \epsilon \;\dot p_\mu,\; \qquad \;\;\;
\delta_r \psi_\mu = \epsilon \;\dot \psi_\mu,\; \nonumber\\
\delta_r \psi_5 &=& \epsilon \;\dot \psi_5,\; \qquad
\delta_r \chi = {\displaystyle \frac{d}{d\tau}}\; \Bigl (\epsilon \chi \Bigr ),
\;\qquad
\delta_r e = {\displaystyle \frac{d}{d\tau}}\; \Bigl (\epsilon e \Bigr ).
\end{array} \eqno(2.2)
$$
It should noted that (i) $\delta_r \Sigma (\tau) = \Sigma^\prime (\tau)
- \Sigma (\tau)$ for the generic field variable $\Sigma = x_\mu, p_\mu, e,
\psi_\mu, \psi_5, \chi$, and (ii) the gauge fields $e$ and $\chi$ do
transform in a similar fashion (and distinctly different from the rest of the
field variables). The first- and the second-order Lagrangians 
are  endowed with the first-class constraints $\Pi_e \approx 0, 
\Pi_\chi \approx 0, p^2 - m^2 \approx 0, p \cdot \psi - m \psi_5 \approx 0$ 
in the language of Dirac's prescription for the classification of
constraints. Here $\Pi_e$ and $\Pi_\chi$ are the canonical conjugate
momenta corresponding to the einbein field $e(\tau)$ and the fermionic
gauge field $\chi$, respectively. There are second-class constraints 
too in the theory but we shall {\it not} concentrate on them for 
our present discussion. The existence of
the first-class constraints $\Pi_e \approx 0$ and
$p^2 - m^2 \approx 0$ on this physical system, generates the following
gauge symmetry transformation $\delta_g$ for the field variables of 
the first-order Lagrangian $L_f^{(m)}$ 
(for the description of a spinning relativistic particle):
$$
\begin{array}{lcl}
\delta_g x_\mu &=& \xi\; p_\mu,\; \qquad\;
\delta_g p_\mu = 0,\; \;\qquad\;
\delta_g \psi_\mu = 0, \nonumber\\
\delta_g \psi_5 &=& 0,\; \;\;\qquad\;\;\;
\delta_g \chi = 0,\;
\;\;\qquad\;\;\;
\delta_g e = \dot \xi,
\end{array} \eqno(2.3)
$$
where $\xi (\tau)$ is an infinitesimal gauge parameter. The pair of
fermionic constraints $\pi_\chi \approx 0$ and $p \cdot \psi - m \psi_5
\approx 0$, generate the following supergauge symmetry transformations 
$\delta_{sg}$
for the bosonic and fermionic field variables of the first-order
Lagrangian $L_f^{(m)}$:
$$
\begin{array}{lcl}
\delta_{sg} \;x_\mu &=& \kappa\; \psi_\mu,\; \;\;\qquad\;
\delta_{sg} \;p_\mu = 0,\; \;\;\qquad\;
\delta_{sg} \;\psi_\mu = i\; \kappa \;p_\mu, \nonumber\\
\delta_{sg} \;\psi_5 &=& i \;\kappa\; m,\; \;\;\qquad\;
\delta_{sg} \;\chi = i \;\dot \kappa,\;\;\qquad\;
\delta_{sg} \;e = 2\;\kappa\; \chi,
\end{array} \eqno(2.4)
$$
where $\kappa (\tau)$ is an infinitesimal fermionic
(i.e. $\kappa^2 = 0$) supergauge transformation
parameter. The above infinitesimal 
transformations are {\it symmetry} transformations because:
$$
\begin{array}{lcl}
\delta_{r} L^{(m)}_f &=& {\displaystyle \frac{d}{d\tau}}\;
\Bigl [\; \epsilon\; L^{(m)}_f\; \Bigr ],\;
\qquad
\delta_{g} L^{(m)}_f = {\displaystyle \frac{d}{d\tau}\;
\Bigl [\; \frac{\xi}{2}} \;\Bigl (p^2 + m^2 \Bigr)\; \Bigr ], 
\nonumber\\
\delta_{sg}\; L^{(m)}_f &=& {\displaystyle \frac{d}{d\tau}\;
\Bigl [\; \frac{\kappa}{2}} \;
\Bigl (p \cdot \psi + m \; \psi_5 \Bigr)\; \Bigr ]. 
\end{array} \eqno(2.5)
$$
It is straightforward to check that, for the generic field variable $\Sigma$,
we have  $(\delta_g - i \delta_{sg}) \Sigma = \delta_r \Sigma$ with the
identifications $\xi = e \epsilon$ and $\kappa = \chi \epsilon$ and
validity of the on-shell conditions (i.e. $\dot p_\mu = 0,\;
\dot \psi_\mu = \chi\;p_\mu,\; \dot \psi_5 = \chi\;m, \;\dot x_\mu
= e\;p_\mu - i\;\chi\;\psi_\mu,\; p^2 = m^2,\; p \cdot \psi = m \psi_5$).

The gauge- and supergauge symmetry transformations (2.3) and (2.4) can
be combined together and generalized to the nilpotent 
BRST symmetry transformations. The usual trick of the BRST prescription
could be exploited here to express the gauge- and supergauge parameters
$\xi = \eta c$ and $\kappa = \eta \beta$ in terms of the fermionic
($c^2 = 0$) and bosonic ($\beta^2 \neq 0$) ghost fields 
and $\eta$. It will be noted that  $\eta$ is
the spacetime independent anticommuting (i.e. $\eta c + c \eta = 0$, etc.)
parameter which is required to maintain the bosonic nature
of $\xi$ (in $\xi = \eta c$) and the fermionic nature of $\kappa$ (in $\kappa
= \eta \beta$). 
The ensuing nilpotent ($(s^{(0)}_b)^2 = 0$) BRST transformations [29],
for the spinning relativistic particle,   are
\footnote{We follow here the notations and conventions adopted in
[30,31]. In its full blaze of glory, the true nilpotent 
(anti-)BRST transformations $\delta_{(A)B}$ are the product of an 
(anticommuting) spacetime
independent parameter $\eta$ and the nilpotent transformations $s_{(a)b}$.
It is clear that $\eta$ commutes with all the bosonic (even) fields of the
theory and anticommutes with fermionic (odd) fields
(i.e. $\eta c + c \eta = 0, \eta \bar c + \bar c \eta = 0$, etc.).}
$$
\begin{array}{lcl}
s^{(0)}_b x_\mu &=& c p_\mu + \beta \psi_\mu,\;
\qquad s^{(0)}_b c =  - i \;\beta^2,\; 
\qquad s^{(0)}_b p_\mu = 0,\; \qquad 
s^{(0)}_b \psi_\mu = i \beta p_\mu, \nonumber\\
s^{(0)}_b \bar c &=& i\; b,\; \;\;\qquad s^{(0)}_b b = 0,\; \;\;\qquad 
s^{(0)}_b e =  \dot c + 2 \;\beta \;\chi,\; \;\;\;\qquad 
s^{(0)}_b \chi = i\; \dot \beta,\; \nonumber\\
s^{(0)}_{b} \psi_5 &=& i\; \beta \;m,\; \;\;\;\qquad 
s^{(0)}_{b} \beta = 0,\; \;\;\;\qquad \;
s^{(0)}_{b} \bar \beta = i \;\gamma,\; \;\;\;\qquad\;
s^{(0)}_{b} \gamma = 0.
\end{array} \eqno(2.6)
$$
The above off-shell nilpotent ($(s^{(0)}_b)^2 = 0$)
transformations are the {\it symmetry}
transformations for the system
because the following Lagrangian (which is the generalization
of $L_f^{(m)}$):
$$
\begin{array}{lcl}
 L^{(m)}_{b} &=&   p \cdot \dot x - 
{\displaystyle \frac{1}{2}}\;e\; (p^2 - m^2) 
+ {\displaystyle \frac{i}{2} \; \Bigl ( \psi \cdot \dot \psi
- \psi_5 \;\dot \psi_5 \Bigr )  
+ i\; \chi\; \Bigl (p \cdot \psi  - \psi_5 m \Bigr )} \nonumber\\
&+& b\; \dot e + \gamma \;\dot \chi 
+ {\displaystyle \frac{1}{2}}\; b^2 - i\; \dot {\bar c}\; \Bigl (\dot c
+ 2\; \chi\; \beta \Bigr ) - \dot {\bar \beta}\; \dot \beta,
\end{array} \eqno(2.7)
$$
transforms to a total derivative under (2.6), namely;
$$
\begin{array}{lcl}
s^{(0)}_b L^{(m)}_b = {\displaystyle \frac{d} {d \tau} \;
\Bigl [ \;\frac{1}{2}\; c\; (p^2 + m^2) 
+ \frac{1}{2}\; \beta \; (p \cdot \psi + m \psi_5)
+ b\; (\dot c + 2 \beta \chi) - i \gamma \dot \beta} \;\Bigr ].
\end{array} \eqno(2.8)
$$
A few comments are in order now. First, the first-order Lagrangian $L_f^{(m)}$
has been extended to include the gauge-fixing term and 
the Faddeev-Popov ghost terms (constructed by the bosonic as well as
the fermionic (anti-)ghost fields) in $L_b^{(m)}$. Second, the 
bosonic auxiliary field $b$ and
the fermionic auxiliary field $\gamma$ (with $\gamma^2 = 0, \gamma \chi
+ \chi \gamma = 0, c \gamma + \gamma c = 0$, etc.)
are the Nakanishi-Lautrup fields. Third, the fermionic (i.e. ${\bar c}^2  = 0,
c \bar c + \bar c c = 0,$ etc.) anti-ghost field $\bar c$ and the
bosonic (i.e. $\bar \beta^2 \neq 0, \beta \bar \beta = \bar \beta \beta$, etc.)
anti-ghost field $\bar\beta$ are required in the theory to have a precise
nilpotent BRST symmetry for the system under consideration. Fourth, the
above nilpotent transformations (2.6) are generated by the conserved
($\dot {Q^{(0)}_b} = 0$) and nilpotent (i.e. $(Q^{(0)}_b)^2 = 0$) BRST charge
$Q^{(0)}_b$ as given below:
$$
\begin{array}{lcl}
Q^{(0)}_b = {\displaystyle \frac{c}{2}\; \Bigl (p^2 - m^2 \Bigr ) + \beta\;
\Bigl (p \cdot \psi - m\; \psi_5 \Bigr ) + b\; \Bigl (\dot c + 2 \;\beta\;\chi
\Bigr ) + \dot {\bar c}\; \beta^2 - i \gamma\;\dot \beta}. 
\end{array} \eqno(2.9)
$$
Fifth, the nilpotent ($(s^{(0)}_{ab})^2 = 0$)
anti-BRST symmetry transformations $s^{(0)}_{ab}$ and corresponding generator 
$Q^{(0)}_{ab}$ can be computed from (2.6) and (2.9) by the substitutions:
$c \leftrightarrow \bar c$ and $\beta \leftrightarrow \bar \beta$. Sixth,
the above generators and corresponding symmetries obey 
the property of anticommutativity
(i.e. $s^{(0)}_b s^{(0)}_{ab} + s^{(0)}_{ab} s^{(0)}_b = 0, 
Q^{(0)}_b Q^{(0)}_{ab} 
+ Q^{(0)}_{ab} Q^{(0)}_b = 0$). Finally, the conservation of the BRST charge
$Q^{(0)}_b$ can be proven by exploiting the equations of motion
$$
\begin{array}{lcl}
&&\dot p_\mu = 0,\; \qquad \dot x_\mu = e p_\mu - i \chi \psi_\mu,\; \qquad
\dot \psi_\mu = \chi p_\mu,\; \qquad \dot \psi_5 = \chi m,\; \qquad 
\dot \chi = 0, \nonumber\\
&& b = - \dot e,\; \qquad \dot b = - \frac{1}{2}\; (p^2 - m^2),\;
\qquad \ddot \beta = 0,\;
\qquad \ddot {\bar \beta} = 2 i \dot {\bar c} \chi, 
\qquad \ddot {\bar c} = 0,\;
\nonumber\\
&& \ddot c + 2\; \dot \beta\; \chi + 2 \;\beta\; \dot \chi = 0,\; \qquad
\dot \gamma + 2\; i\; \dot {\bar c}\; \beta 
+ i\; (p \cdot \psi - m \psi_5) = 0,
\end{array} \eqno(2.10)
$$
derived from the BRST invariant Lagrangian $L^{(m)}_b$ of (2.7).

For our further discussion, 
we deal with the limiting cases of (2.6) and (2.7) so that
we can study the BRST transformations corresponding to 
the gauge transformations (2.3) and the supergauge transformations
(2.4), separately and independently. 
It is evident that $\beta \to 0, \bar\beta \to 0,
\gamma \to 0$ in (2.6) leads to the {\it nilpotent} 
(($ s^{(1)}_b)^2 = 0$) BRST transformations $s^{(1)}_b$, 
corresponding to the gauge transformations (2.3), as
$$
\begin{array}{lcl}
s^{(1)}_{b} x_\mu &=& c\; p_\mu,\; \qquad\;\;
s^{(1)}_b p_\mu = 0,\; \qquad\; s_b^{(1)} c = 0,\; \qquad\;
s^{(1)}_b \psi_\mu = 0, \nonumber\\
s^{(1)}_b \psi_5 &=& 0,\; \quad s^{(1)}_b \bar c = i b,\; 
\quad s^{(1)}_b b = 0,\;
\quad s^{(1)}_b \chi = 0,\; \qquad s^{(1)}_b  e = \dot c, 
\end{array} \eqno(2.11)
$$
which are the symmetry transformations for the Lagrangian
$$
\begin{array}{lcl}
 L^{(1)}_{b} &=&   p \cdot \dot x - 
{\displaystyle \frac{1}{2}}\;e\; (p^2 - m^2) 
+ {\displaystyle \frac{i}{2}} \; \Bigl ( \psi \cdot \dot \psi
- \psi_5 \;\dot \psi_5 \Bigr ) \nonumber\\ 
&+& i\; \chi\; \Bigl (p \cdot \psi  - \psi_5 m \Bigr ) 
+ b\; \dot e 
+ {\displaystyle \frac{1}{2}}\; b^2 - i\; \dot {\bar c}\; \dot c,
\end{array} \eqno(2.12)
$$
obtained from (2.7) under the above specific limits
(i.e. $(\beta, \bar \beta, \gamma) \to 0$). In another limiting case
(i.e. $c \to 0, \bar c \to 0, b \to 0$) of (2.6), we obtain the following
{\it non-nilpotent} ($(s^{(2)}_b)^2 \neq 0$)
 BRST transformations, corresponding
to the supergauge transformations (2.4), as 
$$
\begin{array}{lcl}
s^{(2)}_b x_\mu &=& \beta\; \psi_\mu,\; \qquad\;
s^{(2)}_b p_\mu = 0,\; \qquad\;\; s^{(2)}_b \beta = 0,\;\; \qquad
s^{(2)}_b \psi_\mu = i\; \beta \;p_\mu, \nonumber\\
s^{(2)}_b \psi_5 &=& i \;\beta\; m,\; \quad
s^{(2)}_b \chi = i \;\dot \beta,\;
\quad s^{(2)}_b \bar \beta = i \gamma,\; \quad s^{(2)}_b \gamma = 0,\; \quad
s^{(2)}_b e = 2\;\beta\; \chi,
\end{array} \eqno(2.13)
$$
that are found to be the symmetry transformations for the Lagrangian
$$
\begin{array}{lcl}
 L^{(2)}_{b} &=&   p \cdot \dot x - 
{\displaystyle \frac{1}{2}}\;e\; (p^2 - m^2) 
+ {\displaystyle \frac{i}{2} \; \Bigl ( \psi \cdot \dot \psi
- \psi_5 \;\dot \psi_5 \Bigr )}  \nonumber\\
&+& i \;\chi\; \Bigl (p \cdot \psi  - \psi_5 m \Bigr ) 
+ \gamma\; \dot \chi - \dot {\bar \beta}\; \dot \beta, 
\end{array} \eqno(2.14)
$$
derived from (2.7) under the above limits: $(c, \bar c, b) \to 0$.
It will be noted that (i) the anti-BRST versions of (2.11) and (2.13) can
be obtained by the substitutions $c \leftrightarrow \bar c$ and
$\beta \leftrightarrow \bar\beta$, respectively. (ii) In a similar
fashion, the generators $Q^{(1,2)}_b$
for $s^{(1,2)}_b$, can be derived from (2.9)
by taking into account the above limiting cases. (iii) The equations of motion
for the Lagrangians (2.12) and (2.14) can be derived from (2.10) under the
limits cited above.

We wrap up this section with a couple of comments on the {\it nilpotency}
property of the BRST transformations $s^{(2)}_b$ in (2.13) that correspond
to the supergauge transformations in (2.4). First, it is straightforward to
note that, under the restriction $\beta^2 = 0$, one can restore the nilpotency
(i.e. $(s^{(2)}_b)^2 = 0$) for the transformations $s^{(2)}_b$. This
aspect of $\beta$ can be fulfilled if this bosonic ghost field $\beta$
is taken to be a composite (i.e. $\beta \sim c_1 c_2$) of a couple of
fermionic ($c_1^2 = c_2^2 = 0, c_1 c_2 + c_2 c_1 = 0$) ghost fields
$c_1$ and $c_2$ [29]. Second, if this condition (i.e. $\beta^2 = 0$) is true,
it turns out that the two BRST symmetry transformations $s^{(1,2)}_b$
really decouple from each-other in the sense that $\{s^{(1)}_b, s^{(2)}_b \}
= 0$. In such a situation, they can distinctly be separate from each-other
as well as from $s_b$ in (2.6) (with $(s^{(1,2)}_b)^2 = 0, 
\{s^{(1)}_b, s^{(2)}_b \} = 0$) [29]. It is, ultimately,
interesting to note that {\it all} the nilpotent symmetry transformations
$s^{(i)}_r$ (with $i = 0, 1, 2$), for the generic field 
$\Sigma$ of the system,
can be succinctly expressed in terms of the conserved and nilpotent 
charges $Q^{(i)}_r$ (with $i = 0, 1, 2$) as

$$
\begin{array}{lcl}
s^{(i)}_r \Sigma = - i\; [\Sigma, Q^{(i)}_r]_{\pm},\; \qquad
r = b, ab,\; \qquad \Sigma = x_\mu, p_\mu, b, e, c, \bar c, \chi, \gamma,
\beta, \bar \beta, \psi_\mu, \psi_5.
\end{array} \eqno(2.15)
$$
The $\pm$ signs, 
present as the subscripts on the above square
brackets, stand for the brackets to be 
(anti-)commutators for the generic field $\Sigma$
being (fermionic)bosonic in nature.\\

\noindent
{\bf 3 Gauge BRST Symmetries: Augmented Superfield Approach}\\

\noindent
To derive the nilpotent ($(s^{(1)}_{(a)b})^2 = 0$)
(anti-)BRST transformations $s^{(1)}_{(a)b}$ (cf. (2.11)) for
the gauge (einbein) field $e(\tau)$
and the (anti-)ghost fields $(\bar c)c$ in the superfield formalism (where
the horizontality condition plays a decisive role),
we begin with a 
general three ($1 + 2$)-dimensional supermanifold
parametrized by the superspace coordinates $Z = (\tau, \theta, \bar \theta)$
where $\tau$ is an
even (bosonic) coordinate
and $\theta$ and $\bar \theta$ are the two odd (Grassmannian) coordinates
(with $\theta^2 = \bar \theta^2 = 0, 
\theta \bar \theta + \bar \theta \theta = 0)$. On this supermanifold, one can
define a 1-form supervector superfield $\tilde V = d Z (\tilde A)$ with
$\tilde A (\tau,\theta,\bar\theta) = (E (\tau, \theta, \bar \theta), 
\;F (\tau, \theta, \bar \theta), \;\bar F (\tau, \theta, \bar \theta))$
as the component multiplet superfields. The superfields $E, F, \bar F $ 
can be expanded in terms of the basic fields ($e, c, \bar c$) and auxiliary 
field ($b$) along with  some extra secondary fields 
(i.e. $f, \bar f, B, g, \bar g, s, \bar s, \bar b$), as given below
(see, e.g., [27,11,12]):
$$
\begin{array}{lcl}
E\; (\tau, \theta, \bar \theta) &=& e (\tau) 
\;+\; \theta\; \bar f  (\tau) \;+\; \bar \theta\; f (\tau) 
\;+\; i \;\theta \;\bar \theta \;B (\tau), \nonumber\\
F\; (\tau, \theta, \bar \theta) &=& c (\tau) 
\;+\; i\; \theta\; \bar b (\tau)
\;+\; i \;\bar \theta\; g (\tau) 
\;+\; i\; \theta\; \bar \theta \;s (\tau), \nonumber\\
\bar F\; (\tau, \theta, \bar \theta) &=& \bar c (\tau) 
\;+\; i \;\theta\;\bar g (\tau)\; +\; i\; \bar \theta \;b (\tau) 
\;+\; i \;\theta \;\bar \theta \;\bar s (\tau).
\end{array} \eqno(3.1)
$$
It is straightforward to note that the local 
fields $ f (\tau), \bar f (\tau),
c (\tau), \bar c (\tau), s (\tau), \bar s (\tau)$ on the r.h.s.
are fermionic (anti-commuting) 
in nature and the bosonic (commuting) local fields in (3.1)
are: $e (\tau), B (\tau), g (\tau), \bar g (\tau),
b (\tau), \bar b (\tau)$. It is evident
that, in the above expansion, the bosonic-
 and fermionic degrees of freedom match. This requirement is essential
for the validity and sanctity of any arbitrary supersymmetric theory in the 
superfield formulation. In fact, all the secondary fields will be expressed 
in terms of basic fields due to the restrictions emerging from the application 
of the horizontality condition (see, e.g., [11,12,27])
$$
\begin{array}{lcl} 
\tilde d\; \tilde V = d\; A  = 0,\;\; \qquad d = d \tau \; \partial_\tau,
\; \qquad \; A = d \tau\; e(\tau),\;\; \qquad d^2 = 0,
\end{array} \eqno(3.2)
$$
where the super exterior derivative $\tilde d$ and 
the super connection  1-form $\tilde V$ are defined as
$$
\begin{array}{lcl}
\tilde d &=&  d \tau\; \partial_\tau\;
+ \;d \theta \;\partial_{\theta}\; + \;d \bar \theta \;\partial_{\bar \theta},
\nonumber\\
\tilde V &=& d \tau \;E (\tau , \theta, \bar \theta)
+ d \theta\; \bar F (\tau, \theta, \bar \theta) + d \bar \theta\;
F (\tau, \theta, \bar \theta).
\end{array}\eqno(3.3)
$$
It will be noted that super 1-form 
connection $\tilde V$ is overall bosonic in nature
because of the fact that the superfields $E$ and ($F, \bar F$) are bosonic
($ E^2 \neq 0$) 
and fermionic ($F^2 = \bar F^2 = 0$), respectively. They have been 
combined together with $d\tau$ and
($d \theta, d \bar\theta$) in such a fashion that 
$\tilde V$ becomes bosonic.
We expand $\tilde d \;\tilde V$, present in the l.h.s. of (3.2), as
$$
\begin{array}{lcl}
\tilde d \;\tilde V  &=& 
(d \tau \wedge d \theta) (\partial_{\tau} \bar F - \partial_{\theta} E) 
- (d \theta \wedge d \theta)\; (\partial_{\theta}
\bar F) + (d \tau \wedge d \bar \theta)
(\partial_{\tau} F - \partial_{\bar \theta} E), \nonumber\\
&-& (d \theta \wedge d \bar \theta) (\partial_{\theta} F
+ \partial_{\bar \theta} \bar F) 
- (d \bar \theta \wedge d \bar \theta)
\;(\partial_{\bar\theta} F).
\end{array}\eqno (3.4)
$$
Ultimately, the application of the horizontality condition
($\tilde d \tilde V = d A = 0$) yields
$$
\begin{array}{lcl}
f \;(\tau) &=& \dot c (\tau),\;\; \qquad \;\;
\bar f\; (\tau) = \dot {\bar c} (\tau),\;\; 
\;\;\;\qquad \;\;\; s\; (x) = \bar s\; (x) = 0,
\nonumber\\
B\; (\tau) &=& \dot b (\tau), \;\qquad\;\;\;
b\; (\tau) + \bar b \;(\tau) = 0,\;\; \qquad \;\;\;
g\; (\tau)  = \bar g (\tau) = 0.
\end{array} \eqno(3.5)
$$
It may be emphasized 
that the gauge (einbein) field $e (\tau)$ is a
scalar potential depending only on a single 
variable parameter $\tau$. This is why the 
curvature is zero (i.e. $d A = 0$) because $d\tau \wedge d \tau = 0$
\footnote{It is interesting to point out that, unlike the above case, 
for the 1-form ($A = dx^\mu A_\mu$)
Abelian gauge theory, where the gauge field is a vector potential $A_\mu (x)$,
the 2-form curvature $d A = \frac{1}{2} (dx^\mu \wedge dx^\nu)\; F_{\mu\nu}$
is not equal to zero and it
defines the field strength tensor $F_{\mu\nu} = \partial_\mu A_\nu 
- \partial_\nu A_\mu$ for the Abelian gauge theory. For the 1-form non-Abelian
gauge theory, the Maurer-Cartan equation $F = d A + A \wedge A$ defines
the 2-form $F$ which, in turn, leads to the derivation of the corresponding
group valued field strength tensor $F_{\mu\nu}$.}. 
The insertion of all the above values (cf. (3.5))
in the expansion (3.1) leads to
the derivation of the (anti-)BRST symmetry transformations  for the 
gauge- and (anti-)ghost fields of the theory. This 
statement can be expressed, in an explicit form, as given below:
$$
\begin{array}{lcl}
E\; (\tau, \theta, \bar \theta) &=& e (\tau) 
+ \theta\; \dot {\bar c }(\tau) + \bar \theta\;  \dot c (\tau)) 
+ i \;\theta \;\bar \theta \;\dot b (\tau),\; \nonumber\\
F \; (\tau, \theta, \bar \theta) &=& c (\tau) 
- i\; \theta\;  b (\tau), \nonumber\\
\bar F\; (\tau, \theta, \bar \theta) &=& \bar c (\tau) 
+ i\; \bar \theta \;b (\tau).
\end{array} \eqno(3.6)
$$
In addition, this exercise provides  the physical interpretation for the
(anti-)BRST charges $Q^{(1)}_{(a)b}$ 
as the generators (cf. (2.15)) of translations 
(i.e. $ \mbox{Lim}_{\bar\theta \rightarrow 0} (\partial/\partial \theta),
 \mbox{Lim}_{\theta \rightarrow 0} (\partial/\partial \bar\theta)$)
along the Grassmannian
directions of the supermanifold. Both these observations can be succinctly 
expressed, in a combined way, by re-writing the super expansion (3.1) as
\footnote{ It is worthwhile to note that the anti-BRST transformations 
$s^{(1)}_{ab}$ for the system, described by the Lagrangian in (2.12), 
are: $s^{(1)}_{ab} x_\mu = \bar c p_\mu,
s^{(1)}_{ab} \bar c = 0, s^{(1)}_{ab} p_\mu = 0, s^{(1)}_{ab} c = - i b, 
s^{(1)}_{ab} b = 0, s^{(1)}_{ab} \chi = 0,  s^{(1)}_{ab} \psi_5 = 0,  
s^{(1)}_{ab} \psi_\mu = 0, s^{(1)}_{ab} e = \dot {\bar c}$. The key point,
that should be emphasized, is the minus sign in : $s^{(1)}_{ab} c = - i b$.}  
$$
\begin{array}{lcl}
E\; (\tau, \theta, \bar \theta) &=& e (\tau) 
+ \;\theta\; (s^{(1)}_{ab} e(\tau)) 
+ \;\bar \theta\; (s^{(1)}_{b} e(\tau)) 
+ \;\theta \;\bar \theta \;(s^{(1)}_{b} s^{(1)}_{ab} e (\tau)), \nonumber\\
F \; (\tau, \theta, \bar \theta) &=& c (\tau) \;
+ \; \theta\; (s^{(1)}_{ab} c (\tau))
\;+ \;\bar \theta\; (s^{(1)}_{b} c (\tau)) 
\;+ \;\theta \;\bar \theta \;(s^{(1)}_{b}\; s^{(1)}_{ab}  (\tau)),
\nonumber\\
\bar F\; (\tau, \theta, \bar \theta) &=& \bar c (\tau) 
\;+ \;\theta\;(s^{(1)}_{ab} \bar c (\tau)) \;
+\bar \theta\; (s^{(1)}_{b} \bar c (\tau))
\;+\;\theta\;\bar \theta \;(s^{(1)}_{b} \;s^{(1)}_{ab} \bar c (\tau)).
\end{array} \eqno(3.7)
$$
It should be noted that the third and fourth terms of the expansion
for $F$ and the second and fourth terms of the expansion for $\bar F$
are zero because ($s^{(1)}_b c = 0, \; s^{(1)}_{ab} \bar c = 0$).

Let us concentrate on the derivation of the 
nilpotent (i.e. $(s^{(1)}_{(a)b})^2 = 0$)
(anti-)BRST symmetry transformations $s^{(1)}_{(a)b}$ 
for the Lorentz scalar fields $\chi (\tau), \psi_5 (\tau)$ and the Lorentz
vector target fields $(x_\mu (\tau), \psi_\mu (\tau), p_\mu (\tau))$. In the
derivation of the nilpotent transformations 
for these fields, under the framework of 
augmented superfield formalism, it is the 
invariance of the conserved quantities  on the supermanifold
that plays a key role. However, at times, one has to tap the inputs from
the super expansions (3.6), derived after the application of the horizontality
condition, for the precise
derivations of the nilpotent transformations. 
Thus, to be very precise, it is the interplay of the horizontality
condition and the invariance of the conserved charges that enables us to 
derive the nilpotent (anti-)BRST transformations.
To justify this
assertion, first of all, we start off with the super expansion of the
superfields $(X^\mu, P_\mu)(\tau, \theta,\bar\theta)$),
corresponding to the ordinary target variables $(x^\mu, p_\mu)(\tau)$
(that specify the Minkowski cotangent manifold and are present 
in the first-order Lagrangian $L^{(m)}_f$), as
$$
\begin{array}{lcl} 
X_\mu (\tau, \theta, \bar\theta) &=& x_\mu (\tau)
\;+ \;\theta\; \bar R_\mu (\tau)\; + \;\bar \theta \; R_\mu (\tau) 
\;+\; i \;\theta \;\bar \theta \; S_\mu (\tau),
\nonumber\\
P_\mu (\tau, \theta, \bar\theta) &=& p_\mu (\tau)
\;+\; \theta \;\bar F_\mu (\tau) \;+ \;\bar \theta \; F_\mu (\tau) 
\;+\; i\; \theta \;\bar \theta \; T_\mu (\tau).
\end{array} \eqno(3.8)
$$
It is evident that, in the limit 
$(\theta, \bar\theta) \rightarrow 0$,
we get back the canonically conjugate
target space variables $(x^\mu (\tau), p_\mu (\tau))$ of the 
first-order Lagrangian $L^{(m)}_f$ in (2.1). Furthermore, the number of
bosonic fields ($x_\mu, p_\mu, S_\mu, T_\mu)$ do match with the fermionic
fields $(F_\mu, \bar F_\mu, R_\mu, \bar R_\mu)$ 
so that the above expansion becomes consistent
with the basic tenets of supersymmetry. All the component fields
on the r.h.s. of the expansion (3.8) are functions of the
monotonically increasing parameter $\tau$ of the world-line. As emphasized in
Section 2, three most decisive features of the free relativistic particle
are (i) $ \dot p_\mu = 0$, 
(ii) $p \cdot \psi - m \psi_5 = 0$,
and (ii) $ p^2 - m^2 = 0$. To be very specific, it can be seen
that the conserved {\it gauge} charge $Q_{g} = \frac{1}{2} (p^2 - m^2)$
couples to the `gauge' (einbein) field $e(\tau)$ in the Lagrangian
$L^{(m)}_f$ to maintain the local gauge invariance 
\footnote{Exactly the same kind of gauge coupling exists 
between the Dirac fields for the fermions (electrons, positrons, quarks, etc.)
and the gauge boson field of the {\it interacting}
1-form (non-)Abelian gauge theories where the matter conserved current
$J_\mu = \bar\psi \gamma_\mu \psi$, constructed by the Dirac fields,
couples to the gauge field $A_\mu$ of the (non-)Abelian
gauge theories to maintain the local gauge invariance (see, e.g., [32]).}
under the transformations (2.3). For the BRST invariant
Lagrangian (2.12), the same kind of coupling exists for
the local BRST invariance to be maintained
in the theory. The invariance of the mass-shell
condition $(p^2 - m^2 = 0$) (i.e. a conserved 
and gauge invariant quantity) as well as 
the conservation of the gauge invariant momenta ($\dot p_\mu = 0$) on the 
supermanifold, namely;
$$
\begin{array}{lcl}
P_\mu (\tau,\theta, \bar\theta) P^\mu (\tau, \theta, \bar\theta) 
- m^2 = p_\mu (\tau) p^\mu (\tau) - m^2,\; \qquad \dot P_\mu (\tau, \theta,
\bar \theta) = \dot p_\mu (\tau),
\end{array} \eqno(3.9)
$$
imply the following restrictions:
$$
\begin{array}{lcl}
F_\mu (\tau) = \bar F_\mu (\tau) = T_\mu (\tau) = 0, \;\;\mbox{and}\;\;
P_\mu (\tau, \theta, \bar\theta) = p_\mu (\tau). 
\end{array} \eqno(3.10)
$$
In other words, the
invariance of the mass-shell condition as well as the conserved momenta
on the supermanifold
enforces $P_\mu (\tau, \theta, \bar\theta)$ to be independent of the
Grassmannian variables $\theta$ and $\bar\theta$.
To be consistent with our earlier interpretations for the (anti-)BRST charges,
in the language of translation generators along the Grassmannian directions
$(\theta)\bar\theta$
of the supermanifold, it can be seen that the above equation can be 
re-expressed as
$$
\begin{array}{lcl}
P_\mu (\tau, \theta, \bar\theta) = p_\mu (\tau) 
+ \theta \; (s^{(1)}_{ab} p_\mu (\tau))
+\bar \theta \; (s^{(1)}_{b} p_\mu (\tau))
+ \theta \; \bar\theta\;(s^{(1)}_b s^{(1)}_{ab} p_\mu (\tau)).
\end{array}\eqno (3.11)
$$
The above equation, {\it vis-{\`a}-vis} (3.10),
makes it clear that $s^{(1)}_b p_\mu (\tau) = 0$
and $s^{(1)}_{ab} p_\mu (\tau) = 0$.

Before we shall derive the nilpotent (anti-)BRST transformations for
$x_\mu (\tau)$, it is useful to compute these transformations
for the fermionic gauge field $\chi (\tau)$ and other fields
$\psi_5 (\tau)$ as well as $ \psi_\mu (\tau)$. To this end in mind,
let us have the super expansions for the superfields, corresponding
to these fields, as follows
$$
\begin{array}{lcl}
K\;\; (\tau,\theta,\bar\theta) &=& \chi (\tau) + \theta\; \bar b_1 (\tau)
+ \bar\theta\; b_1 (\tau) + \theta\; \bar\theta\; f_1 (\tau), \nonumber\\
\Psi_5 \; (\tau,\theta,\bar\theta) &=& \psi_5 (\tau) + \theta\; \bar B_5 (\tau)
+ \bar\theta\; B_5 (\tau) + \theta\; \bar\theta\; f_5 (\tau), \nonumber\\
\Psi_\mu \; (\tau,\theta,\bar\theta) 
&=& \psi_\mu (\tau) + \theta\; \bar b_\mu (\tau)
+ \bar\theta\; b_\mu (\tau) + \theta\; \bar\theta\; f_\mu (\tau).
\end{array} \eqno(3.12)
$$
It will be noted that, in the limit ($\theta, \bar\theta) \to 0$, we do obtain
the usual local fields $\chi (\tau), \psi_5 (\tau)$ and $\psi_\mu (\tau)$
and the fermionic ($\chi, \psi_5, \psi_\mu, f_\mu, f_1, f_5$) and
bosonic ($b_1, \bar b_1, B_5, \bar B_5, b_\mu, \bar b_\mu$) degrees of
freedom do match in the above expansion. 
Let us focus on the conserved quantities $\dot \chi = 0, \ddot \psi_5 = \dot
\chi m = 0, \ddot \psi_\mu = \dot \chi p_\mu + \chi \dot p_\mu = 0$
(cf. (2.10)). The invariance of $\dot \chi (\tau) = 0$ on the 
supermanifold
\footnote{This condition, in a more sophisticated language, is just
the gauge choice for the system.}
 leads to the following consequences:
$$
\begin{array}{lcl}
\dot K (\tau, \theta, \bar \theta) = \dot \chi (\tau) = 0 \qquad
\Rightarrow \qquad \dot b_1 = \dot {\bar b_1} = \dot f_1 = 0.
\end{array}\eqno (3.13)
$$
One of the solutions is $b_1 = \bar b_1 = f_1 = C$ where $C$ is a 
$\tau$-independent constant. Geometrically, this amounts to the
shift of the superfield $K (\tau,\theta,\bar\theta)$ along
$\theta$- and $\bar \theta$-directions by a constant value $C$. One
can choose the $\theta$ and $\bar\theta$ axes on the supermanifold
in such a manner that this constant $C$ is zero. 
Thus, ultimately,
we obtain $b_1 = \bar b_1 = f_1 = 0$. Interpreted in the light
of (3.6), (3.7) and (3.11), this shows that $s^{(1)}_{(a)b} \chi = 0$ (i.e.
$K (\tau, \theta, \bar\theta) = \chi (\tau)$).
As a side remark, it is worthwhile to mention that
the above explicit equality (3.13) can be re-expressed as the equality
of the conserved quantities {\it only}. In other words,
the restriction $K (\tau,\theta,\bar\theta) = \chi (\tau)$,
directly implies that $b_1 = \bar b_1 = f_1 = 0$. For the (non-)Abelian gauge
theories, such kind of equality has been taken into account [24-26]
where only the expression for the conserved quantity has been equated
on the supermanifold. In view 
of the above, it can be seen that the following invariances
of the conserved quantities (cf. (2.10) for details)
on the supermanifold: 
$$
\begin{array}{lcl}
&& \dot \Psi_5 (\tau, \theta, \bar \theta) -  K (\tau, \theta, \bar\theta)\; m
= \dot \psi_5 (\tau) - \chi (\tau) \; m, \nonumber\\
&& \dot \Psi_\mu (\tau, \theta, \bar\theta) -
K (\tau, \theta, \bar \theta) P_\mu (\tau, \theta, \bar\theta) =
\dot \psi_\mu (\tau) - \chi (\tau) p_\mu (\tau),
\end{array}\eqno (3.14)
$$
imply the following restrictions for the expansion in (3.12), namely;
$$
\begin{array}{lcl}
\bar B_5 = B_5  = f_5  = 0,\; \;\;\qquad\;\;
\bar b_\mu = b_\mu  = f_\mu  = 0.
\end{array} \eqno(3.15)
$$
In the above derivation, we have exploited the results of (3.10)
(i.e. $P_\mu (\tau, \theta, \bar\theta) = p_\mu (\tau)$) and
(3.13) (i.e. $K (\tau, \theta, \bar \theta) = \chi (\tau))$. Insertions
of (3.15) into (3.12) imply that $s^{(1)}_{(a)b} \psi_5 = 0$ 
(i.e. $\Psi_5 (\tau,\theta,\bar\theta) = \psi_5 (\tau)$) and
$s^{(1)}_{(a)b} \psi_\mu  = 0$ (i.e. $\Psi_\mu (\tau,\theta,\bar\theta)
= \psi_\mu (\tau)$). 
It is worthwhile to emphasize that
the above solutions are one set of the {\it simplest} solutions which
are of interest to us. 
A more general solution (than the above) might exist.

Now the stage is set for the derivation of the nilpotent (anti-)BRST
transformations for the target space coordinate variable $x_\mu (\tau)$.
One of the most important 
relations, that plays a pivotal role in the derivation of the mass-shell 
condition ($p^2 - m^2 = 0$) for the Lagrangian $L_s^{(m)}$, is
$\dot x_\mu (\tau) = e (\tau) p_\mu (\tau) - i \chi \psi_\mu$. 
This is due to the fact that $(\partial L^{(m)}_s / \partial e) = 0$ implies
that $e^2 m^2 = (\dot x_\mu + i \chi \psi_\mu)^2$ and $p_\mu =
(\partial L^{(m)}_s/ \partial \dot x^\mu) = e^{-1} (\dot x_\mu 
+ i \chi \psi_\mu)$. A simple way
to derive the (anti-)BRST  transformations for the 
coordinate target variable $x_\mu (\tau)$  is to
require the invariance of this
central relation on the supermanifold, as  
$$
\begin{array}{lcl}
&& \dot X_\mu (\tau,\theta,\bar\theta) - E (\tau, \theta, \bar\theta)
P_\mu (\tau, \theta, \bar\theta) + i\; K (\tau, \theta, \bar\theta)
\;\Psi_\mu (\tau, \theta, \bar\theta) \nonumber\\ 
&=& \dot x_\mu (\tau) - e (\tau) \;p_\mu (\tau)
+ i\; \chi (\tau)\; \psi_\mu (\tau),
\end{array} \eqno(3.16)
$$
where $E (\tau,\theta, \bar\theta)$ is the expansion in (3.6) which has been
obtained after the application of the horizontality condition. Exploiting
the relations 
$P_\mu (\tau, \theta, \bar\theta) = p_\mu (\tau)$ from (3.10)
and $K (\tau, \theta, \bar \theta) = \chi (\tau)$, it
can be seen that  the following relations emerge from (3.16):
$$
\begin{array}{lcl}
\dot {\bar R_\mu} = \dot {\bar c}\; p_\mu,\;\; 
\qquad \dot R_\mu = \dot c\; p_\mu,\;\; \qquad \dot S_\mu = \dot b\; p_\mu.
\end{array}\eqno (3.17)
$$
At this crucial stage,  we summon one of the most decisive
physical insights into the characteristic features of a free 
spinning relativistic 
particle which states that there is no action of any kind of
force (i.e. $\dot p_\mu (\tau) = 0$) on the {\it free} motion of the particle. 
Having taken into account this decisive input, we obtain from (3.17),
the following relations
$$
\begin{array}{lcl}
\dot {\bar R}_\mu \equiv
\partial_\tau {\bar R_\mu} = \partial_\tau (\bar c p_\mu),\;\; \qquad
\dot R_\mu \equiv
\partial_\tau R_\mu = \partial_\tau (c p_\mu),\;\; \qquad 
\dot S_\mu \equiv \partial_\tau
 S_\mu = \partial_\tau (b p_\mu),
\end{array}\eqno (3.18)
$$
which lead to
$$
\begin{array}{lcl}
{\bar R_\mu} (\tau) = \bar c\; p_\mu,\;\; \qquad
R_\mu (\tau) = c \;p_\mu,\;\; \qquad 
 S_\mu (\tau) = b \;p_\mu.
\end{array}\eqno (3.19)
$$
The insertions of these values into
 the expansion (3.8) lead to the derivation 
of the nilpotent (anti-)BRST transformations ($s^{(1)}_{(a)b}$) 
on the target space
coordinate field $x_\mu (\tau)$, as
$$
\begin{array}{lcl}
X_\mu (\tau, \theta, \bar\theta) &=& x_\mu (\tau)
+ \;\theta\; (s^{(1)}_{ab} x_\mu(\tau)) + \;\bar \theta \; 
(s^{(1)}_b x_\mu (\tau)) 
+ \;\theta \;\bar \theta \; (s^{(1)}_b s^{(1)}_{ab} x_\mu (\tau)).
\end{array}\eqno (3.20)
$$
In our recent papers [24-26] on interacting 1-form (non-)Abelian gauge 
theories, it has been shown that there is a beautiful
consistency and complementarity between the
horizontality condition and the requirement of the invariance of conserved
matter (super)currents on the supermanifold. The former restriction leads to
the derivation of nilpotent symmetries for the gauge- and (anti-)ghost fields.
The latter restriction yields such 
kind of transformations for the matter fields. For
the case of the 
free spinning relativistic particle, it can be seen that the invariance of
the gauge invariant and conserved quantities on the supermanifold, leads to
the derivation of the transformations for the target field
variables. To corroborate
this assertion, we see that the conserved and gauge invariant charge
$Q_g = \frac{1}{2}\; (p^2 - m^2)$ is the analogue of the conserved {\it matter}
current of the 1-form interacting (non-)Abelian gauge theory. Since
the expansion for $P_\mu (x.\theta,\bar\theta)$ is trivial (cf. (3.10)), we
have to re-express the mass-shell condition 
(i.e. $ e^2 (p^2 - m^2)= 
(\dot x_\mu + i \chi \psi_\mu)^2  - e^2  m^2$) in the language
of the superfields (3.6) and $\Psi_\mu (\tau,\theta,\bar\theta)
= \psi_\mu (\tau), K (\tau,\theta,\bar\theta) = \chi (\tau)$. 
Thus, the invariance 
of the conserved (super)charges on the supermanifold is:
$$
\begin{array}{lcl}
&&[\dot X_\mu (\tau,\theta,\bar\theta) + i\; K (\tau, \theta, \bar\theta)\;
\Psi_\mu (\tau, \theta, \bar\theta)]\;
[\dot X^\mu (\tau,\theta,\bar\theta) + i\; K(\tau, \theta, \bar\theta)\;
\Psi^\mu  (\tau, \theta, \bar\theta)]\; \nonumber\\
&& - m^2\;E(\tau,\theta,\bar\theta) E(\tau,\theta,\bar\theta) 
\nonumber\\
&& = [\dot x_\mu (\tau) + i \;\chi (\tau) \; \psi_\mu (\tau)]\;
[\dot x^\mu (\tau) + i\; \chi (\tau) \; \psi^\mu (\tau)]
- m^2\; e^2.
\end{array}\eqno (3.21)
$$
The equality of the appropriate terms from the l.h.s. and r.h.s. leads to 
$$
\begin{array}{lcl}
&& (\dot x_\mu + i\; \chi \;\psi_\mu)\; \dot {\bar R^\mu} 
= m^2\;e\; \dot {\bar c},\;\; \qquad
(\dot x_\mu + i\; \chi\; \psi_\mu)\;\dot R^\mu = m^2\;e\; \dot c,\;\; \qquad
\nonumber\\
&&(\dot x_\mu + i\;\chi\;\psi_\mu)\; \dot S^\mu = m^2\;e\; \dot b,\;\; \qquad
\dot R_\mu \dot {\bar R^\mu} = m^2\;\dot c\; \dot {\bar c}.
\end{array}\eqno (3.22)
$$
Taking the help of the key relation $\dot x_\mu 
+ i\; \chi\; \psi_\mu = e\;p_\mu$, we obtain
the expressions for $\dot R_\mu, \dot {\bar R}_\mu, \dot S_\mu$ exactly same
as the ones given in (3.17) {\it for the mass-shell condition $p^2 - m^2 = 0$ 
to be valid}. Exploiting the no force (i.e. $\dot p_\mu = 0$) criterion
on the free motion of a 
spinning relativistic particle, we obtain the expressions for 
$R_\mu, \bar R_\mu, S_\mu$ in exactly the same form as given in (3.19). 
The insertion of these 
values in (3.8) leads to the same expansion as given in (3.20). This provides 
the geometrical interpretation for the (anti-)BRST charges as the 
{\it translational generators}. 
It should be noted that the
restrictions in (3.16) and (3.21) are intertwined. However, the latter
is more physical because it states the invariance of the conserved and
gauge invariant 
{\it mass-shell} condition {\it explicitly}. As a side remark, we would
like to comment on the other conserved quantity $p \cdot \psi - m \psi_5
= 0$. It is straightforward to check that this conserved quantity is
automatically satisfied on the supermanifold due to the fact that
$\Psi_5 (\tau,\theta,\bar\theta) = \psi_5 (\tau), 
\Psi_\mu (\tau,\theta,\bar\theta) = \psi_\mu (\tau), 
P_\mu (\tau,\theta,\bar\theta) =  p_\mu (\tau)$. In fact, the results
$\Psi_5 (\tau,\theta,\bar\theta) = \psi_5 (\tau), 
\Psi_\mu (\tau,\theta,\bar\theta) = \psi_\mu (\tau) $ can be 
obtained from this very conserved quantity if we follow the same trick
as we have exploited, in the above, for the conserved and gauge invariant
quantity
$p^2 - m^2 = 0$ for the derivation of $s^{(1)}_{(a)b}$ for $x_\mu (\tau)$.\\

\noindent
{\bf 4 Supergauge BRST Symmetries: Augmented Superfield Formalism}\\

\noindent
We derive here the nilpotent ($ (s^{(2)}_{(a)b})^2 = 0$)
(anti-)BRST symmetry transformations $s^{(2)}_{(a)b}$ (cf. (2.13)),
corresponding to the supergauge transformations in (2.4),
in the framework of augmented superfield formalism.
The crucial assumption here is the condition $\beta^2 = 0$
which can be satisfied if and only if the bosonic (anti-)ghost fields
$(\bar\beta)\beta$ were made up of two fermionic ghost fields.
In contrast to the super 1-form {\it bosonic} connection $\tilde V$, quoted
in (3.3), we define here a fermionic super 1-form connection $\tilde {\cal F}$
as follows: 
$$
\begin{array}{lcl}
\tilde {\cal F} = d Z (\tilde F)
= d \tau\; K (\tau, \theta, \bar\theta) 
+ i\; d \theta\;\bar {\cal B} (\tau, \theta, \bar\theta)
+ i\; d \bar\theta\;{\cal B} (\tau, \theta, \bar\theta),
\end{array}\eqno (4.1)
$$
where the supermultiplet $\tilde F (\tau, \theta, \bar\theta)
= (K (\tau,\theta,\bar\theta), i\; {\cal B} (\tau,\theta,\bar\theta), 
i\;\bar {\cal B} (\tau, \theta, \bar\theta)$) has three
superfields and the fermionic super gauge 
field $K (\tau, \theta, \bar\theta)$ has an expansion as given in (3.12).
The bosonic superfields ${\cal B} (\tau,\theta, \bar\theta)$ and
$\bar {\cal B} (\tau,\theta,\bar\theta)$ have the following expansions
$$
\begin{array}{lcl}
{\cal B}\; (\tau,\theta,\bar\theta) &=& 
\beta (\tau) + 
i\;\theta\; \bar f_2 (\tau)
+ i\;\bar\theta\; f_3 (\tau) + i\; \theta\; \bar\theta\; b_2 (\tau), \nonumber\\
\bar {\cal B} \; (\tau,\theta,\bar\theta) &=& 
\bar \beta (\tau) + 
i\;\theta\; \bar f_3 (\tau)
+ i\;\bar\theta\; f_2 (\tau) + i\; \theta\; \bar\theta\; \bar b_2 (\tau),
\end{array} \eqno(4.2)
$$
which yield the bosonic (anti-)ghost fields $(\bar\beta)\beta$ in the limit
$(\theta, \bar\theta) \to 0$ and the bosonic 
($\beta, \bar\beta, b_2, \bar b_2$) 
and fermionic ($f_2, \bar f_2, f_3, \bar f_3$) degrees of freedom
do match. The requirement of the horizontality condition (with the super
exterior derivative $\tilde d$ defined in (3.3)):
$$
\begin{array}{lcl} 
\tilde d\; \tilde {\cal F} = d\; {\cal F}  = 0,\;\; \qquad \;\;
d = d \tau \; \partial_\tau,\; \qquad
{\cal F} = d \tau\; \chi (\tau),\; \qquad d^2 = 0,
\end{array} \eqno(4.3)
$$
leads to the derivation of the secondary fields in terms of the basic fields
as well as the auxiliary fields. In fact, the explicit expression for
$\tilde d \tilde {\cal F}$ is:
$$
\begin{array}{lcl}
\tilde d \;\tilde {\cal F}  &=& 
(d \tau \wedge d \theta) (i\; \partial_{\tau} \bar {\cal B}
 - \partial_{\theta} K) 
- i\; (d \theta \wedge d \theta)\; (\partial_{\theta}
\bar {\cal B}) + (d \tau \wedge d \bar \theta)
(i\;\partial_{\tau} {\cal B} - \partial_{\bar \theta} K) \nonumber\\
&-& i\; (d \theta \wedge d \bar \theta) (\partial_{\theta} {\cal B}
+ \partial_{\bar \theta} \bar {\cal B}) 
- i\; (d \bar \theta \wedge d \bar \theta)
(\partial_{\bar \theta} {\cal B}). 
\end{array}\eqno(4.4)
$$
The application of the horizontality condition $\tilde d \tilde {\cal F}
= d {\cal F} = 0$, leads to the following relations:
$$
\begin{array}{lcl}
&& \partial_\theta \; \bar {\cal B} = 0,\; \qquad
 \partial_{\bar \theta} \;  {\cal B} = 0,\; \qquad
 \partial_{\bar\theta} \; \bar {\cal B} +
 \partial_{\theta} \;  {\cal B} = 0, \nonumber\\
&& \partial_\theta\; K = i \;\partial_\tau\; \bar {\cal B},\;\; \qquad \;\;
\;\;\; \partial_{\bar\theta}\; K = i \;\partial_\tau\;  {\cal B}. 
\end{array}\eqno (4.5)
$$
The first two relations, in the above, produce $b_2 = \bar b_2 = 0, f_3 =
\bar f_3 = 0$. The third one leads to $f_2 + \bar f_2 = 0$. Choosing
$ f_2 = \gamma$ implies that $\bar f_2 = - \gamma$ and the bosonic
superfields ${\cal B}$ and $\bar {\cal B}$ become chiral- and anti-chiral
superfields, respectively, with the following expansions: 
$$
\begin{array}{lcl}
&&{\cal B} (\tau, \theta) = \beta (\tau) - i\; \theta\; \gamma
\equiv \beta (\tau) - \theta\; (s^{(2)}_{ab} \beta (\tau)), \nonumber\\
&&\bar {\cal B} (\tau, \bar\theta) = \bar \beta (\tau) + i\; \theta\; \gamma
\equiv \bar \beta (\tau) + \bar\theta\; (s^{(2)}_{b} \bar \beta (\tau)).
\end{array}\eqno (4.6)
$$
It will be noted that the Nakanishi-Lautrup fermionic ($\gamma^2 = 0$)
auxiliary field $\gamma (\tau)$
is {\it not} a basic dynamical field variable of the
theory (cf. (2.14)).
The above equations establish that $s^{(2)}_b \beta (\tau) = 0$ and
$s^{(2)}_{ab} \bar \beta (\tau) = 0$.
Exploiting the expressions for ${\cal B} (\tau, \theta)$ and
$\bar {\cal B} (\tau, \bar\theta)$ (cf. (4.6)) 
in the last two relations of (4.5),
we obtain the following values for the component local
secondary fields of the expansion
for  $K (\tau,\theta,\bar\theta)$ in (3.12):
$$
\begin{array}{lcl}
b_1 (\tau) = i \; \dot \beta (\tau),\;
\qquad \;\bar b_1 (\tau) = i\; \dot {\bar\beta} (\tau),\; \qquad\;
f_1 (\tau) = - \dot \gamma (\tau). 
\end{array}\eqno (4.7)
$$
The insertions of the above values in the expansion of 
$K (\tau,\theta,\bar\theta)$ yields:
$$
\begin{array}{lcl}
K\; (\tau,\theta,\bar\theta) &=& \chi (\tau) + \theta\; 
(s^{(2)}_{ab} \chi (\tau))
+ \bar\theta\; (s^{(2)}_b \chi (\tau)) 
+ \theta\; \bar\theta\; (s^{(2)}_b\; s^{(2)}_{ab}\; \chi (\tau)).
\end{array}\eqno (4.8)
$$
It will be noted that the expressions in (4.6) for the bosonic
superfields can also be written in an exactly the same form
as (4.8) as $s^{(2)}_b \beta (\tau) = 0$ and 
$s^{(2)}_{ab} \bar \beta (\tau) = 0$. There is only one caveat, however.
This has to do with the $(-)$ sign in the expansion of $B (\tau, \theta)$
\footnote{ It will be noted that the explicit form of 
the anti-BRST transformations $s^{(2)}_{ab}$ for the system are:
$s^{(2)}_{ab} x_\mu = \bar \beta \psi_\mu, s^{(2)}_{ab} p_\mu = 0,
s^{(2)}_{ab} \bar \beta = 0, s^{(2)}_{ab} \beta = + i \gamma,
s^{(2)}_{ab} \gamma = 0, s^{(2)}_{ab} \psi_\mu = i \bar \beta p_\mu,
s^{(2)}_{ab} \chi = i \dot {\bar \beta}, 
s^{(2)}_{ab}  e= 2 {\bar \beta} \chi$. The key point that should be emphasized
is the fact that, the bosonic nature of the
(anti-)ghost fields $(\bar\beta)\beta$
does {\it not} allow a change of sign between $s^{(2)}_{ab} \beta 
(= + i \gamma)$ and $s^{(2)}_b \bar \beta (= + i \gamma)$ for
the symmetry of the Lagrangian to be maintained. 
This situation is totally opposite to the case
of $s^{(1)}_{(a)b}$ (cf. (2.11))
where the (anti-)ghost fields are fermionic in nature,
and, that is why, the change of sign in
$s^{(1)}_{ab} c = - i b, s^{(1)}_b \bar c = i b$ is required.}.
We shall dwell on it, in detail, in the conclusions part (i.e. section 5)
of our present paper.
It is clear that this exercise provides the geometrical interpretation for the
(anti-)BRST charges $Q^{(2)}_{(a)b}$ 
as the generators (cf. (2.15)) of translations 
(i.e. $ \mbox{Lim}_{\bar\theta \rightarrow 0} (\partial/\partial \theta),
 \mbox{Lim}_{\theta \rightarrow 0} (\partial/\partial \bar\theta)$)
along the Grassmannian directions $(\theta)\bar\theta$
of the three $(1 + 2)$-dimensional supermanifold.

Let us focus on the derivation of the nilpotent transformations
$s^{(2)}_{(a)b}$ for the target field variables ($x_\mu (\tau), p_\mu (\tau),
\psi_\mu (\tau)$) and the Lorentz scalar fermionic field $\psi_5 (\tau)$
in the framework of augmented superfield formalism. Here, once again, the
interplay of the horizontality condition and the
invariance of the conserved quantities on the supermanifold do play
a very important and decisive roles. To see it clearly, 
let us first concentrate on
the invariance of the conserved quantities (given in (3.14)) on
the supermanifold. The explicit substitutions 
of super expansions yield the following relationships:
$$
\begin{array}{lcl}
&& (\dot \psi_5 + \theta\; \dot {\bar B}_5 + \bar \theta\; \dot B_5
+ \theta\;\bar\theta\;\dot f_5) -
(\chi + i \;\theta\;\dot {\bar \beta} + i\; \bar\theta\; \dot \beta
- \theta\; \bar\theta\; \dot \gamma)\; m = \dot \psi_5 - \chi\;m,
\nonumber\\
&& (\dot \psi_\mu + \theta\; \dot {\bar b_\mu} + \bar \theta\; \dot b\mu
+ \;\theta\;\bar\theta\;\dot f_\mu) -
(\chi + i\;\theta\;\dot {\bar \beta} + i\; \bar\theta\; \dot \beta
- \theta\; \bar\theta\; \dot \gamma)\; p_\mu = \dot \psi_\mu - \chi\;p_\mu,
\end{array}\eqno (4.9)
$$
where we have exploited the expansions of 
$\Psi_\mu (\tau,\theta,\bar\theta)$ and
$\Psi_5 (\tau,\theta,\bar\theta)$  given in (3.12) and have used the
expansion of $K (\tau,\theta,\bar\theta)$ from (4.8) that has been 
obtained after the application of the horizontality condition. It is
evident that the following relations emerge between the
secondary component fields and basic fields
$$
\begin{array}{lcl}
&& \dot B_5 = 
i\;\dot \beta\; m \equiv  \partial_\tau (i\;\beta \;m),\; \;\qquad\;
 \dot {\bar B_5} = i\;\dot{\bar \beta}\; m \equiv  \partial_\tau (i\;
\bar\beta \;m),  \nonumber\\
&& 
 \dot f_5 = - \dot \gamma\; m \equiv - \partial_\tau (\gamma \;m),\; 
\;\qquad\;
 \dot {\bar b_\mu} = i\;\dot{\bar \beta}\; p_\mu \equiv  
\partial_\tau (i\;\bar\beta \;p_\mu), \nonumber\\
&& \dot b_\mu =
i\;\dot \beta\; p_\mu \equiv  \partial_\tau (i\;\beta \;p_\mu),
\;\qquad\;
 \dot f_\mu = - \dot \gamma\; p_\mu \equiv - \partial_\tau (\gamma \;p_\mu), 
\end{array}\eqno (4.10)
$$
where, in the latter set of entries, we have used the requirement of the
free motion ($\dot p_\mu = 0$) of a free spinning relativistic particle.
Ultimately, the insertions of the above values in the expansions (3.12),
yields the following expansions in terms of $s^{(2)}_{(a)b}$ 
(cf. (2.13)):
$$
\begin{array}{lcl}
\Psi_5\; (\tau,\theta,\bar\theta) &=& \psi_5 (\tau) + \theta\; 
(s^{(2)}_{ab} \psi_5 (\tau))
+ \bar\theta\; (s^{(2)}_b \psi_5 (\tau)) 
+ \theta\; \bar\theta\; (s^{(2)}_b\; s^{(2)}_{ab}\; \psi_5 (\tau)),
\nonumber\\
\Psi_\mu\; (\tau,\theta,\bar\theta) &=& \psi_\mu (\tau) + \theta\; 
(s^{(2)}_{ab} \psi_\mu (\tau))
+ \bar\theta\; (s^{(2)}_b \psi_\mu (\tau)) 
+ \theta\; \bar\theta\; (s^{(2)}_b\; s^{(2)}_{ab}\; \psi_\mu (\tau)).
\end{array}\eqno (4.11)
$$
The above expansions produce the same geometrical interpretations
for the symmetries  $s^{(2)}_{(a)b}$ and the generators $Q^{(2)}_{(a)b}$ 
(i.e. the translational generators along the Grassmannian directions) as
such conclusions drawn for the expansion in (4.8) 
for $K (\tau,\theta,\bar\theta)$.

Having obtained the super-expansions of superfields 
(i.e. $K, {\cal B}, \bar {\cal B}, \Psi, \Psi_5$) in terms of
the local $\tau$-dependent ordinary basic fields in (4.6), (4.8)
and (4.11), the stage is now set for the derivation of the nilpotent symmetry
transformations for the einbein field $e (\tau)$ and the canonically
conjugate target space field variables $x_\mu (\tau)$ and $p_\mu (\tau)$.
It is clear that $\dot p_\mu = 0$ and the mass-shell condition
$p^2 - m^2 = 0$ are (i) supergauge invariant (i. e. 
$\delta_{sg} p_\mu = 0$), and (ii) conserved quantities. Thus, 
their invariance on the supermanifold, once again, leads to the
same conclusions as illustrated in (3.9) and (3.10). As a consequence,
we have $s^{(2)}_{(a)b} p_\mu = 0$. Now, the central problem is to obtain
the nilpotent transformations for $e(\tau)$ and $x_\mu (\tau)$. In this
connection, it turns out that  
(i) the intertwined relations given in (3.16) and (3.21), and
(ii) the conserved quantity $p \cdot \psi - m \psi_5 = 0$ (expressed
in terms of $p_\mu = e^{-1} \; (\dot x_\mu + i \chi \psi_\mu)$
so that it becomes $ \dot x \cdot \psi -  m \; e\; \psi_5 = 0$), emerge to
help in the final computation. Exploiting the explicit expressions for the
expansions of $\Psi_\mu (\tau,\theta,\bar\theta)$ and 
$\Psi_5 (\tau,\theta,\bar\theta)$ given in (4.11) and inserting the
values of $\dot X_\mu (\tau,\theta,\bar\theta)$ and 
$E (\tau,\theta,\bar\theta)$ from (3.8) and (3.1) in the following
invariance of the conserved quantity on the supermanifold:
$$
\begin{array}{lcl}
\dot X \cdot \Psi - m\; E \; \Psi_5 = \dot x \cdot \psi - m \; e\; \psi_5,
\end{array}\eqno (4.12)
$$
we obtain the following relationships
$$
\begin{array}{lcl}
&& \dot {\bar R} \cdot \psi - m \; \bar f\; \psi_5
= i \; e\; \bar\beta\; m^2 - i\; \bar\beta\; (\dot x \cdot p), \nonumber\\
&& \dot  R \cdot \psi - m \;  f\; \psi_5
= i \; e\; \beta\; m^2 - i\; \beta \; (\dot x \cdot p), \nonumber\\
&& i \; (\dot S \cdot \psi) + i\;\bar\beta\; (\dot R \cdot p) - i \; \beta \;
(\dot {\bar R} \cdot p) + i\; \bar f\;\beta\; m^2 - i\; f \;\bar\beta\; m^2
- i\;B\; \psi_5 \nonumber\\
&&= \gamma\; (\dot x \cdot p) - e\; \gamma\; m^2.
\end{array}\eqno (4.13)
$$
Similarly, tapping the potential of the cute relationship in (3.16)
(which is finally intertwined with the invariance of the conserved
quantity in (3.21)), we obtain the following connections among the
component secondary local fields and the basic local fields:
$$
\begin{array}{lcl}
&& \dot R_\mu - f\;p_\mu = \dot \beta\; \psi_\mu - \beta\; \chi\; p_\mu,
\nonumber\\
&& \dot {\bar R}_\mu - \bar f\;p_\mu 
= \dot {\bar\beta}\; \psi_\mu - {\bar\beta}\; \chi\; p_\mu,
\nonumber\\
&& i\; \dot S_\mu - B\;p_\mu = i\;\dot \gamma\; \psi_\mu 
+ i \; (\dot {\bar \beta}\; \beta - \dot \beta \bar \beta)\;p_\mu
+ i\;  \chi\; \gamma\; p_\mu,
\end{array}\eqno (4.14)
$$
where the explicit expansions from (3.1) (for $E$), (3.8) (for $X_\mu$),
(4.8) (for $K$) and (4.11) (for $\Psi_\mu)$ have been used. Solution to
the above equations on the on-shell (i.e. $\dot \psi_\mu = \chi\; p_\mu,
\dot p_\mu = 0, \dot x_\mu = e \;p_\mu - i \; \chi\;\psi_\mu,
p^2 - m^2 = 0, p \cdot \psi - m \; \psi_5 = 0$, etc.) are as follows: 
$$
\begin{array}{lcl}
&&f =  2\; \beta\; \chi,\; \qquad \;\bar f = 2 \; \bar\beta\; \chi,\;
\qquad\; R_\mu = \beta\; \psi_\mu, \;
\qquad\; \bar R_\mu = \bar \beta\; \psi_\mu, \nonumber\\
&& S_\mu = \gamma\; \psi_\mu + \beta \; \bar \beta\; p_\mu,\; \qquad\;
B = 2 \; \dot \beta\; \bar \beta + 2 \;\gamma\; \chi.
\end{array}\eqno (4.15)
$$
As a side remark, it is interesting to point out that, even from a single
relationship in (4.14), some of the above values could be guessed.
For instance, the relationship 
$\dot R_\mu - f\; p_\mu = \dot \beta \; \psi_\mu
- \beta\; \chi\; p_\mu$ 
can be re-expressed as
$\dot R_\mu - f\; p_\mu = \partial_\tau (\beta \; \psi_\mu) - \beta\;
\dot \psi_\mu - \beta\; \chi\; p_\mu$. 
Exploiting the on-shell 
condition $\dot \psi_\mu = \chi\; p_\mu$ in the above, it can be
seen that 
$\dot R_\mu - f\; p_\mu = \partial_\tau (\beta \; \psi_\mu) - 
2\;\beta\; \chi\; p_\mu$. This last relation gives a glimpse
of $R_\mu = \beta \; \psi_\mu$ and $f = 2\;\beta\;\chi$. In exactly the 
same manner, it can be seen that $\bar R_\mu = \bar \beta\; \psi_\mu$
and $\bar f = 2 \; \bar\beta \; \chi$. With these values,
other expressions of (4.15) follow, which ultimately, satisfy
all the relations derived in  (4.13) and (4.14).
Insertions of the above values in the expressions for superfield
$E (\tau,\theta,\bar\theta)$ in (3.1) and the 
superfield $X_\mu (\tau,\theta,\bar\theta)$
in (3.8) lead to the following:
$$
\begin{array}{lcl}
E\; (\tau, \theta, \bar \theta) &=& e (\tau) 
\;+\; \theta\; (s^{(2)}_{ab} e  (\tau)) \;+\; \bar \theta\; 
(s^{(2)}_b e (\tau)) 
\;+\;\theta \;\bar \theta \;(s^{(2)}_b s^{(2)}_{ab} e (\tau)), \nonumber\\
X_\mu (\tau, \theta, \bar\theta) &=& x_\mu (\tau)
\;+ \;\theta\; (s^{(2)}_{ab} x_\mu (\tau))\; 
+ \;\bar \theta \; (s^{(2)}_b x_\mu (\tau)) 
\;+\;\theta \;\bar \theta \; (s^{(2)}_b s^{(2)}_{ab} x_\mu (\tau)).
\end{array}\eqno (4.16)
$$
The above equation, once again, establishes the fact that the
nilpotent (anti-)BRST generators $Q^{(2)}_{(a)b}$ (and corresponding
symmetry transformations $s^{(2)}_{(a)b}$) are the translational
generators along the Grassmannian directions $(\theta)\bar\theta$ of the
supermanifolds.\\

\noindent
{\bf 5 Conclusions}\\

\noindent
In our present endeavour, we have exploited, in an
elegant way, the key ideas of the augmented
superfield formalism to derive two
sets of anticommuting  (i.e. $\{ s^{(1)}_{(a)b}, s^{(2)}_{(a)b} \} = 0$)
and nilpotent ($(s^{(1,2)}_{(a)b})^2 = 0$) (anti-)BRST symmetry transformations
$s^{(1,2)}_{(a)b}$ for {\it all} the fields variables, present in the Lagrangian
description of a free massive spinning relativistic particle. The
theoretical arsenal of (i) the horizontality condition, and (ii) the
invariance of the conserved quantities on the supermanifold, have played
very decisive roles in the above derivations. One of the central new features
of our present investigation is the application of the augmented superfield
formulation to a supersymmetric system where the bosonic $(\bar\beta)\beta$
and fermionic $(\bar c)c$ (anti-)ghost fields are present together in the
(anti-)BRST invariant Lagrangian describing the free motion
($\dot p_\mu = 0$) of the super-particle. Of course,
for the system under consideration, the reparametrization symmetry invariance
and the gauge symmetry invariance are also present. All these symmetries
are inter-related.

Even though
the pair of bosonic (anti-)ghost fields $(\bar\beta)\beta$ are commutative
in nature (i.e. $\beta \bar\beta = \bar \beta \beta, \beta\; \Sigma
= \Sigma\; \beta, \bar \beta\; \Sigma = \Sigma\; \bar\beta$ for the generic
field $\Sigma = x_\mu, p_\mu, \psi_\mu, \psi_5, e, \chi, c, \bar c, b, 
\gamma$), they are taken to be nilpotent of order two 
($\beta^2 = 0, \bar\beta^2 = 0$) with the assumption that they are made up of
a pair of {\it fermionic} (anti-)ghost fields (i.e. $\beta \sim c_1 c_2,
\bar \beta \sim \bar c_1 \bar c_2, c_1^2 = c_2^2 = 0, c_1 c_2 
+ c_2 c_1 = 0$, etc.). Such an assumption is essential for a couple of 
advantageous reasons. First, the (anti-)BRST
transformations (3.13) (corresponding to the supergauge symmetry transformations
(2.4)) become nilpotent (i.e. $(s^{(2)}_{(a)b})^2 = 0$)
under the above assumption. This can be
explicitly checked for $(s^{(2)}_{(a)b})^2 \; x_\mu (\tau) = 0$
and $(s^{(2)}_{(a)b})^2 \; e (\tau) = 0$ where  the conditions
$\beta^2 = \bar \beta^2 = 0$
and $\partial_\tau (\beta)^2 = \partial_\tau (\bar\beta)^2 = 0$ are
required for the proof of an explicit  nilpotency. Second,
this assumption also allows $s^{(2)}_{(a)b}$ to decouple from the
nilpotent  transformations in (2.6) (where, in some sense, they are hidden)
and the nilpotent transformations (2.11) so that they could become
completely separate and  independent. At this stage,
it is worth emphasizing that there is no such kind of restriction 
(i.e. $\beta^2 \neq 0$) on the bosonic ghost field $\beta$ in the 
nilpotent transformations listed in (2.6).

In our earlier works [23-27], the horizontality
condition was augmented to include  the invariance
of the conserved matter currents/charges on the
supermanifold. In our present endeavour, 
the augmented superfield formalism has been extended to include
the invariance of {\it any} kind of conserved quantities on the
supermanifold and still (i) there is a mutual consistency and
complementarity between the two above types of restrictions. (ii)
The geometrical interpretations for the nilpotent (anti-)BRST charges
$Q^{(1,2)}_{(a)b}$, as the translational generators
$(\mbox {Lim}_{\bar\theta \to 0} (\partial/\partial\theta))
\mbox {Lim}_{\theta \to 0} (\partial/\partial\bar\theta)$ along the
$(\theta)\bar\theta$-directions of the $(D + 2)$-dimensional supermanifold,
remains intact.
(iii) The nilpotency of the (anti-)BRST charges $Q^{(1,2)}_{(a)b}$
is encoded in a couple of successive translations 
(i.e. $(\partial/\partial\theta)^2 = (\partial/\partial\bar\theta)^2 = 0$)
along either of the two Grassmannian directions of the supermanifold. 
(iv) The anticommutativity of the 
nilpotent (anti-)BRST charges $Q^{(1,2)}_{(a)b}$ (and the transformations
they generate) is captured in the relationship
$(\partial/\partial\theta) (\partial/\partial\bar\theta) +
(\partial/\partial\bar\theta) (\partial/\partial\theta) = 0$. Thus,
our present extension of our earlier works [23-27] is a very
natural generalization of the horizontality condition where the
beauty of the geometrical interpretations is not spoiled in any way.

We dwell a bit on the negative sign in (4.6) for the expansion
of the chiral superfield ${\cal B} (\tau,\theta)$. In fact, the bosonic
ghost term $\dot {\bar\beta} \dot \beta$ in (2.7) (or (2.14)) remains 
invariant under $\beta \to \pm \bar\beta, \bar\beta \to \pm \beta$. We 
have taken the $(+)$ sign for our description of 
the anti-BRST transformations $s^{(2)}_{ab}$. However,
one could choose $\beta \to -\bar\beta, \bar \beta \to - \beta$
equally well. In that case, the negative sign in the expansion
of (4.6) disappears. However, under the latter choice, the beautiful
expansions of the superfields $E (\tau,\theta,\bar\theta)$
and $X_\mu (\tau,\theta,\bar\theta)$ in (4.16) get disturbed and
some minus signs crop up in the expansion. This is why, we have opted for the
$(+)$ sign in ($\beta \to \pm \bar\beta, \bar\beta \to \pm \beta$)
and all the transformations, 
listed in the whole body of the present text, are consistent with it.
Geometrically, it seems that the translation of the anti-chiral
superfield $\bar {\cal B} (\tau, \bar\theta)$ along the
$(+ \bar\theta)$ direction of the supermanifold
produces the BRST $s^{(2)}_b$ transformation
for the anti-ghost field $\bar\beta$. However, the anti-BRST
transformation $s^{(2)}_{ab}$ for the ghost field $\beta$
is produced by the translation of the chiral superfield 
${\cal B} (\tau, \theta)$ along the $(- \theta)$ direction of
the supermanifold. This kind of discrepancy appears, perhaps,
because of the peculiar behaviour of these bosonic (anti-)ghost
fields which are commutative in nature
but are restricted to be nilpotent of order two (i.e. $\beta^2 = 0,
\bar\beta^2 = 0$).

In our present investigation, we have not dwelt on
the derivation of the beautiful 
nilpotent symmetries $s^{(0)}_{(a)b}$
(cf. (2.6)) in the framework of augmented superfield formalism
because we cannot add the bosonic and fermionic 1-form super
connections $\tilde V$ and $\tilde {\cal F}$ (defined in (3.3) and
(4.1)) together. However, we strongly believe that
if (i) the action with Lagrangian (2.7) is written in terms of 
the superfields, and (ii) the BRST symmetry transformations (2.6)
 are expressed in terms of the superfield too, we shall be able
to derive these beautiful nilpotent symmetry transformations
(i.e. without any restrictions) in the framework of augmented
superfield approach to BRST formalism. We would like to lay
stress on the fact that
the nilpotent symmetry transformations in (2.6) are beautiful
because there are no restrictions (i.e. $\beta^2 \neq 0,
\bar\beta^2 \neq 0$) and no other peculiarities (like 
its being a composite of two fermions, etc.)
are associated with  the bosonic (anti-)ghost
fields $(\bar\beta)\beta$. Furthermore, our approach
could be extended to be applied to some more complicated
and interesting field theoretic
supersymmetric systems so that, the ideas proposed in our
present endeavour, could be put on a firmer footing. These are
some of the issues that are under investigation and we shall
report these results in our forthcoming publications [33].\\

\noindent
{\bf Acknowledgements}\\

\noindent
The warm hospitality extended by the HEP group of the AS-ICTP, Trieste
is gratefully acknowledged. It is a great pleasure to thank L. Bonora
(SISSA, Trieste), E. Ivanov (JINR, Dubna), K. S. Narain (AS-ICTP)
and G. Thompson (AS-ICTP) for fruitful discussions.

\end{document}